\documentclass{aastex}          
\usepackage{spr-astr-addons}    

\begin{document}
%
\title{Low X-ray Efficiency of a Young High-B Pulsar PSR~J1208$-$6238 Observed with {\it Chandra}}

\shorttitle{PSR~J1208$-$6238 with {\it Chandra}}
\shortauthors{Bamba et al.}


\author{Aya Bamba\altaffilmark{1,2}}
\author{Eri Watanabe\altaffilmark{3}}
\author{Koji Mori\altaffilmark{4}}
\author{Shinpei Shibata\altaffilmark{5}}
\author{Yukikatsu Terada\altaffilmark{6}}
\author{Hidetoshi Sano\altaffilmark{7,8,9}}
\author{Miroslav D. Filipovi\'{c}\altaffilmark{10}}

\altaffiltext{1}{Department of Physics, University of Tokyo, 7-3-1 Hongo, Bunkyo-ku, Tokyo 113-0033, Japan\\
{\tt bamba@phys.s.u-tokyo.ac.jp}}
\email{bamba@phys.s.u-tokyo.ac.jp}
\altaffiltext{2}{Research Center for the Early Universe, School of Science, The University of Tokyo, 7-3-1
Hongo, Bunkyo-ku, Tokyo 113-0033, Japan}
\email{bamba@phys.s.u-tokyo.ac.jp}
\altaffiltext{3}{Department of Physics, Yamagata University, 1-4-12 Kojirakawa-machi, Yamagata, Yamagata 990-8560, Japan}
\altaffiltext{4}{Department of Applied Physics and Electronic Engineering, Faculty of Engineering, University of Miyazaki, 1-1 Gakuen Kibanadai-Nishi, Miyazaki 889-2192, Japan}
\altaffiltext{5}{School of Science and Technology, Yamagata University, 1-4-12 Kojirakawa-machi, Yamagata, Yamagata 990-8560, Japan}
\altaffiltext{6}{Department of Physics, Science, Saitama University, Sakura, Saitama 338-8570, Japan}
\altaffiltext{7}{Institute for Advanced Research, Nagoya University, Furo-cho, Chikusa-ku, Nagoya 464-8601, Japan}
\altaffiltext{8}{Department of Physics, Nagoya University, Furo-cho, Chikusa-ku, Nagoya 464-8601, Japan}
\altaffiltext{9}{National Astronomical Observatory of Japan, 2-21-1 Osawa, Mitaka, Tokyo 181-8588, Japan}
\altaffiltext{10}{Western Sydney University, Locked Bag 1797, Penrith, NSW 2751, Australia}

\begin{abstract}
High magnetic field (high-B) pulsars are key sources
to bridge magnetars and conventional rotation powered pulsars,
and thus
to understand the origin of magnetar activities.
We have estimated a tight upper-limit on the X-ray flux of 
one of the youngest high-B pulsars PSR~J1208$-$6238 for the first time;
a {\it Chandra} 10~ks observation shows 
no significant source.
Depending on the emission models,
the 3$\sigma$ upper-limit on the intrinsic 0.5--7~keV flux
to 2.2--10.0$\times10^{-14}$~erg~s$^{-1}$cm$^{-2}$.
Assuming the distance to the pulsar of 3~kpc,
we suggest that the conversion efficiency from the the spin-down energy to
the X-ray luminosity of this pulsar is almost the smallest among known high-B
pulsars,
and even smaller for those for typical rotation-powered one.
We also discuss possible associations
of a surrounding pulsar wind nebula and a hosting
supernova remnant around this high-B pulsar.
\end{abstract}


\section{Introduction}

It is well known that there are two types of
isolated neutron-star emission-mechanism;
that brings two distinctive populations,
the rotation powered pulsars and the magnetars.
Magnetars are strong X-ray emitters with high variabilities (bursts),
and it was firstly believed that
their properties are caused by the strong dipole magnetic field
$B_d$
(larger than $\sim 10^{13}$~G).
However, several exceptions make the origin unclear,
i.e., the discovery of the low dipole field magnetars,
SGR~0418+5729
(\cite{esposito2010,rea2010}),
and of the rotation powered pulsars with large magnetic
fields comparable to the magnetars (high-B pulsars)
such as PSR~J1119$-$6127
\citep{camilo2000}.
Recent flare detection from this interesting source \citep{archibald2016}
makes that high-B pulsars are key to make clear understanding of 
the origin of magnetar activities.
Another example of magnetar-like bursts on a rotation powered
pulsar would be PSR~J1846$-$0258 \citep{gavriil2008}.
Interestingly, both examples have pulsar wind nebulae (PWNe).
{\it Chandra} resolved PWN tori and jets from
these young pulsar system \citep{ng2004,kargaltsev2008,bamba2010}.
Although the high-B pulsars are now known such objects that bridge the two populations,
the number of the sample is still very small,
and further X-ray observation of high-B pulsars are strongly required.

PSR~J1208$-$6238 was discovered by {\it Fermi}
\citep{clark2016} in the gamma-ray band
on the position of (RA, Dec.)$_{\rm J2000}$ = (12$^{\rm h}$08$^{\rm m}$13$^{\rm s}$.96(6), $-$62$^\circ$38$^{\prime}$002$^{\prime\prime}$.003(4)).
Radio follow-ups detected no signal, implying that this pulsar is a radio-quiet.
The period and period derivative are
0.440590723652(14)~s and 
\\ \noindent
3.2695145(10)$\times 10^{-12}$~s~s$^{-1}$
at 56040 MJD, respectively.
Its young age ($2.14\times 10^3$~yrs),
large spin-down luminosity ($1.5\times 10^{36}$~erg~s$^{-1}$),
and high dipole magnetic field ($3.84\times 10^{13}$~G),
suggest that this pulsar is the third youngest
with the fourth strongest magnetic field,
among 66 pulsas with the dipole magnetic field between $1\times 10^{13}$~G
and $4.4\times 10^{13}$~G.
Actually, these parameters are quite similar to those of 
the youngest and most energetic high-B pulsar, PSR~J1119$-$6127.
The distance is estimated to be 3~kpc, 
under the assumption that the gamma-ray luminosity is proportional to
square root of spin down energy \citep{abdo2013},
which makes this object as one of the nearest known high-B pulsars.
However, no X-ray observation on this source PSR~J1208$-$6238
was performed so far.
Typical rotation-powered pulsars have
X-ray luminosity of the factor of $\sim 10^{-(3-4)}$ of the spin-down energy
\citep{shibata2016},
thus this object should have X-ray luminosity of $\sim 10^{32-33}$~erg~s$^{-1}$.

In this paper, we report on the first X-ray observation result on
PSR~J1208$-$6238.
Section \ref{sec:obs} summarizes the {\it Chandra} observation of this pulsar.
We show the results (\S\ref{sec:results}) and discussion (\S\ref{sec:discuss})
on the pulsar and its surroundings.

\section{Observation and Data reduction}
\label{sec:obs}

{\it Chandra} Advanced CCD Imaging Spectrometer (ACIS)
observed PSR~J1208$-$6238 region on 2019 Dec. 13 in {\tt VFAINT} mode.
We used CIAO 4.11 \citep{fruscione2006} and calibration database 4.8.3
for the data reduction and analysis.
The net exposure was 9.9~ks.
We also used several tools in heasoft 4.11.

\section{Results}
\label{sec:results}

\subsection{PSR~J1208$-$6238}
\label{result:psr}

\begin{figure}
\includegraphics[width=0.47\textwidth]{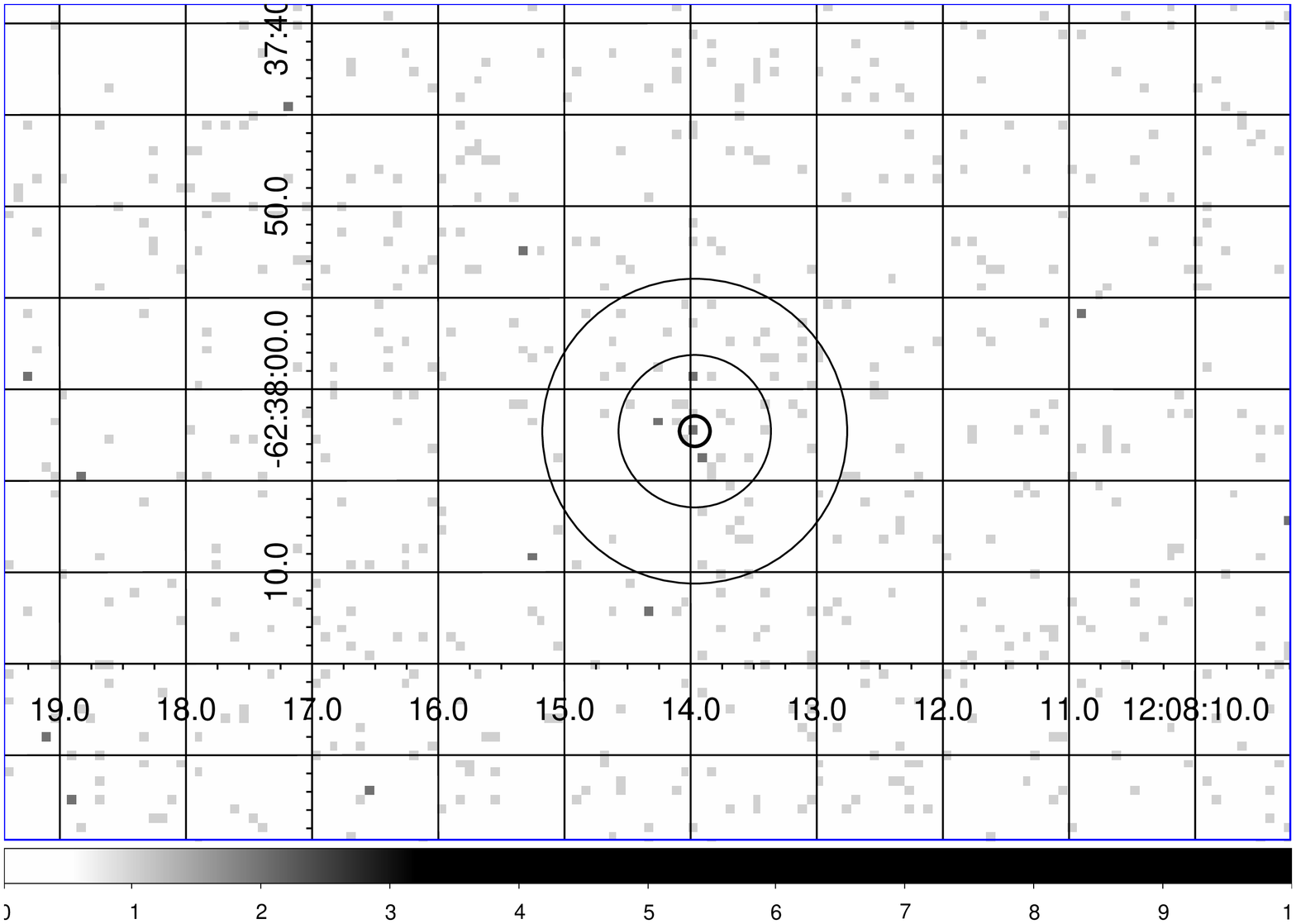}
\caption{All band image of PSR~J1208$-$6238 region.
The thick circle indicates the source region,
whereas the two thin annulus regions indicate the background regions,
respectively,
for the flux estimation.} 
\label{fig:images}
\end{figure}

Figure~\ref{fig:images} shows the all band image
of the
\\ \noindent
PSR~J1208$-$6238 region.
One can see some photons on the position of this pulsar,
but it is not distinct.
The upper-limit was estimated with {\tt srcflux} command in CIAO package
as follows.
We estimated the count rate in the circle of 1.8~pixels,
which is the point spread function radius in this detector position.
The background region was selected from the annuls region
as shown in Figure~\ref{fig:images}.
The estimated count rate is 1.92$\times 10^{-4}$ cps,
which is only 2.7$\sigma$ signifincance.
$3\sigma$ upper-limit is estimated to be 1.0$\times 10^{-3}$~cps.
In order to translate these value to the flux,
we need spectral assumptions.
For the spectral shape of this pulsar,
we assumed two models;
the first one is an absorbed power-law model.
The photon index is assumed to be 2.0,
the same value of that in PSR~J1119$-$6127 \citep{ng2012}.
The other one is an absorbed blackbody,
which is sometimes observed in high-B pulsars and magnetars
\citep{olausen2014,yoneyama2019}.
The temperature is assumed to be 0.2~keV,
which is same as that in PSR~J1119$-$6127
\citep{ng2012}.
For the absorption column in both the emission models,
we have made two assumptions;
the first one is $8\times 10^{21}$~cm$^{-2}$,
from the assumed distance of 3~kpc and the relation
between $N_{\rm H}$ and distance \citep{he2013}.
The other assumption is $N_{\rm H}$ of $1.48\times 10^{22}$~cm$^{-2}$,
which is the total column density to the direction of the pulsar
\citep{HI4PI2016}.
This is the largest (or most pessimistic) assumption.
The resultant absorbed and unabsorbed (intrinsic) upper-limit
with each model is 
summarized in Table~\ref{tab:point-UL}.

\begin{figure}
\includegraphics[width=0.45\textwidth]{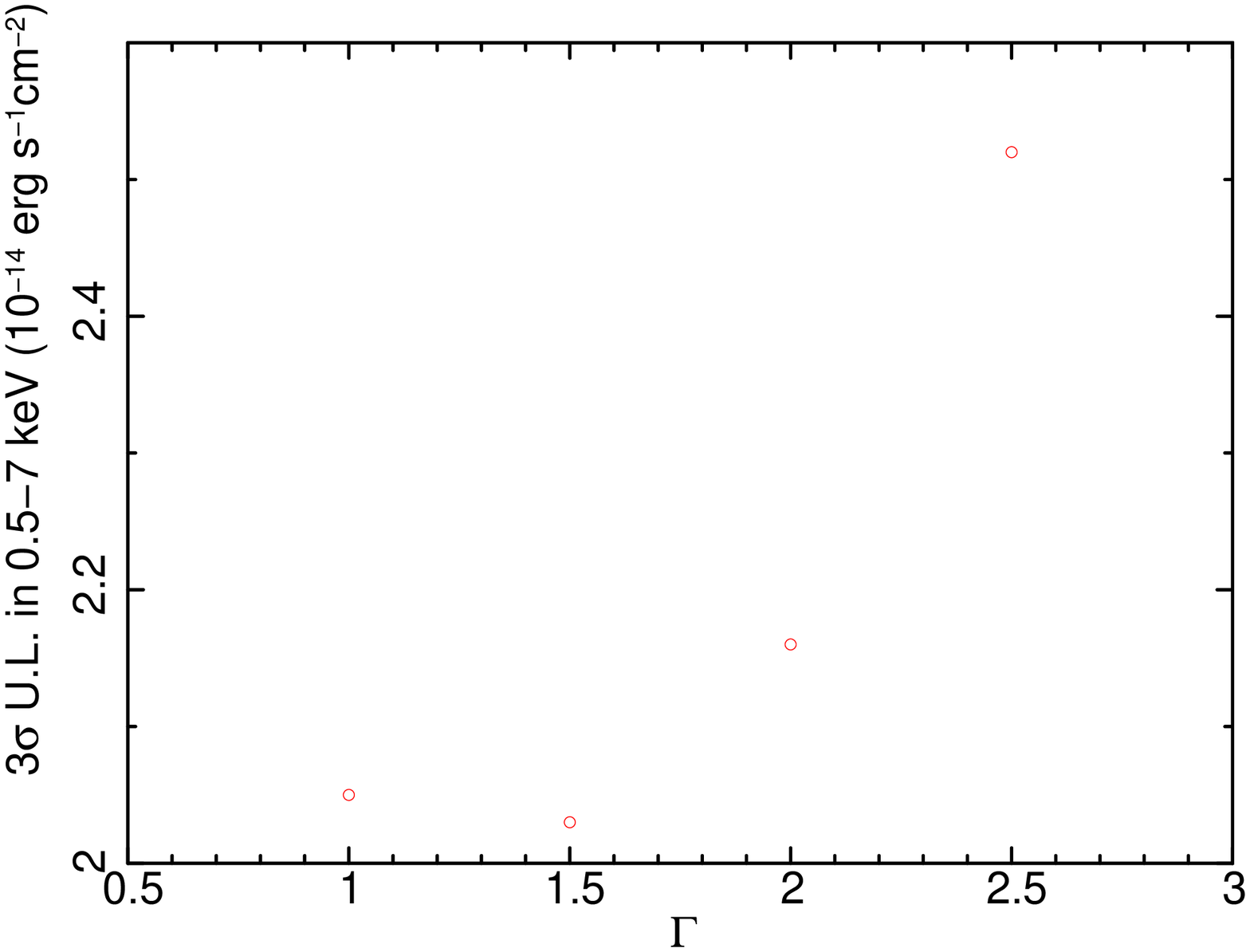}
\includegraphics[width=0.45\textwidth]{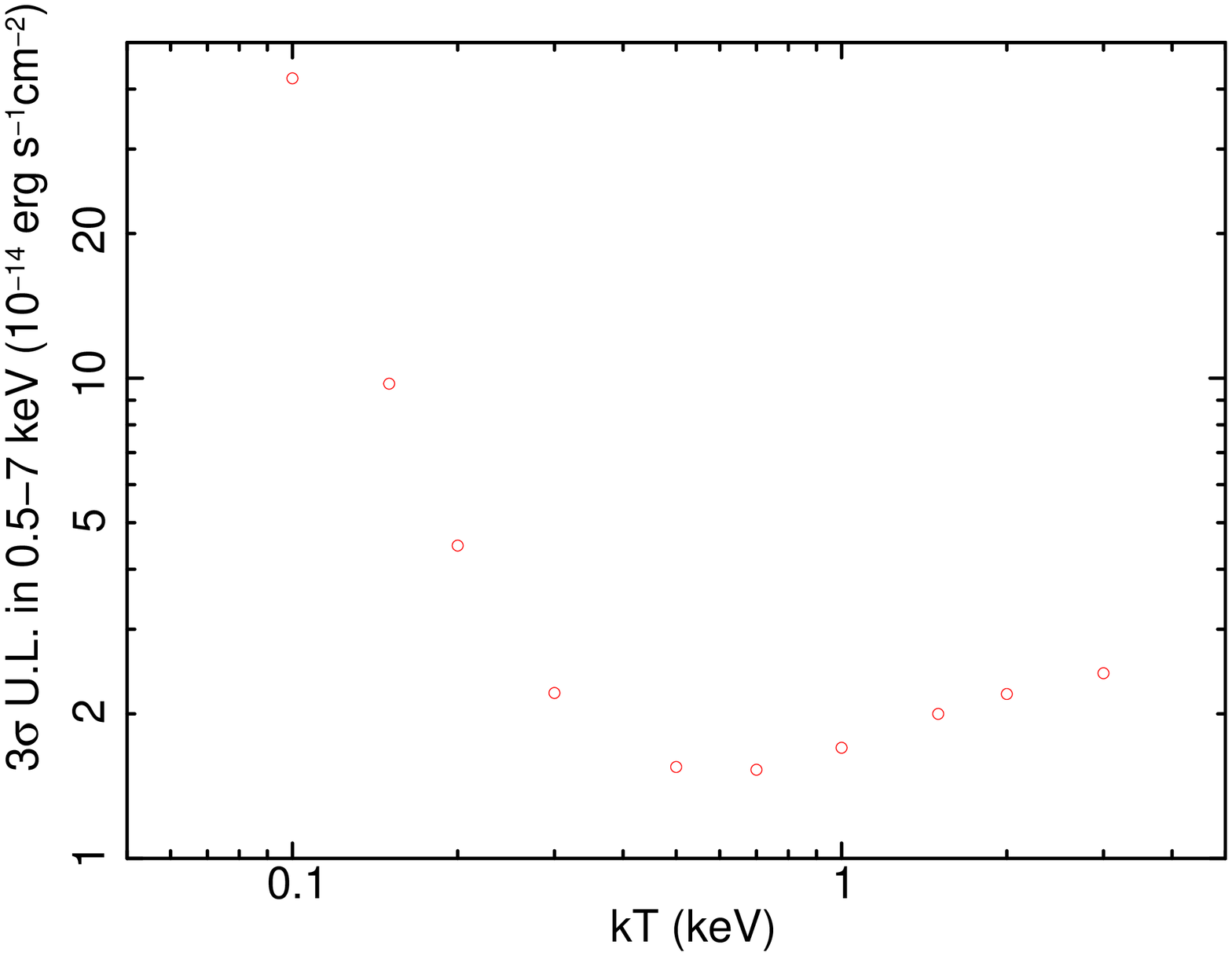}
\caption{3$\sigma$ upper-limit  dependence on
the 0.5--7~keV unabsorbed flux.
The top panel shows the dependence on
the photon index in the power-law model,
whereas the bottom panel shows that on the temperature
in the blackbody model.
The absorption column was fixed to $8\times 10^{21}$~cm$^{-2}$
in the all cases.}
\label{fig:dependence}
\end{figure}

The assumed photon index and temperature also affect the upper-limit.
We checked the 3$\sigma$ upper-limit with the assumed photon index of
1.0, 1.5, 2.0, and 2.5,
and temperature of 0.1, 0.15, 
0.2, 0.3, 0.5, 0.7, 1.0, 1.5, and 2.0~keV.
The absorption column was fixed to be $8\times 10^{21}$~cm$^{-2}$
in the all cases.
The results are summarized in Figure~\ref{fig:dependence}.
In both models,
a softer spectrum results in a larger upper-limit.
This can be due to the stronger coupling with
the absoprtion effect.

\begin{table*}
\caption{3$\sigma$ upper-limit with each model$^\dagger$.}
\label{tab:point-UL}
\begin{center}
\begin{tabular}{p{8pc}cccc}
\hline\hline
Emission model & \multicolumn{2}{c}{$N_{\rm H} = 8\times 10^{21}$~cm$^{-2}$} & \multicolumn{2}{c}{$N_{\rm H} = 1.48\times 10^{22}$~cm$^{-2}$} \\
\hline
& Absorbed & Unabsorbed & Absorbed & Unabsorbed \\
\hline
Power-law model \dotfill & 1.2 & 2.2 & 1.4 & 3.0 \\
Blackbody model \dotfill & 0.6 & 4.5 & 0.6 & 10.0 \\
\hline
\multicolumn{5}{p{30pc}}{$^\dagger$: 0.5--7~keV upper-limit
in the unit of $10^{-14}$~erg~s$^{-1}$cm$^{-2}$.
We assumed $\Gamma=2.0$ for the power-law model,
whereas $kT=0.2$~keV for the blackbody model.
}
\end{tabular}
\end{center}
\end{table*}

Some high-B pulsars show burst-like activities
so the emission can concentrate on some time interval.
We also checked the light curve from the source region.
We have done Kolmogorov-Smirnov test with {\tt lcstats} in headas
and found no significant time variability.
Coherent pulsation search was not done,
with the lack of statistics and timing resolution.

\subsection{Possible PWN}

The region around the pulsar position has several X-ray photons,
which would immediately qualify for the PWN candidate
powered by PSR~J1208$-$6238.
We have made the significance and flux estimation
for the possible diffuse emission.
Here, we treated the background region for the pulsar analysis
shown in Figure~\ref{fig:images} to be the source region.
The background region was newly selected from the source free region
distant from the possible PWN.
The background-subtracted count rate in the 0.5--7.0~keV band was 
estimated with {\tt srcflux} command again,
to be 1.28 (0.70--2.04) $\times 10^{-3}$ cps within 90\% error range.
The significance is calculated to be 4.4$\sigma$ level.
In order to estimate the flux,
we assumed same absorption column to that for PSR~J1208$-$6238
(see \S\ref{result:psr})
and photon index of 2.0.
The estimate absorbed and intrinsic flux in the 0.5--7.0~keV band
is
1.5 (0.8--2.4) $\times 10^{-14}$~erg~cm$^{-2}$s$^{-1}$
and 2.5 (1.4--4.0) $\times 10^{-14}$~erg~cm$^{-2}$s$^{-1}$,
respectively.

\subsection{Counterpart SNR search}

PSR~J1208$-$6238 is allegedly a very young system.
Thus, one would expect to find an associated a young supernova remnant (SNR)
which still emits thermal X-rays.
There is no cataloged SNR in Green Galactic SNR catalog
\citep{green2019}.

\begin{figure}
\includegraphics[width=0.45\textwidth]{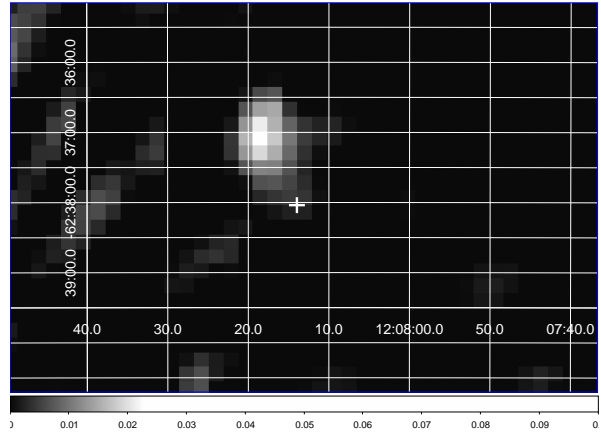}
\caption{SUMSS 843~MHz map around PSR~J1208$-$6238 (cross)
with J2000 coordinates.
The stripe-shape emission on south east is 
the artificial by nearby HII region.
The scale is in linear, with the unit of Jy/beam,
where resolution is 45~arcsec by 45~arcsec.}
\label{fig:radio}
\end{figure}

We searched for the possible radio counterpart using online tools Simbad and 
SkyView\footnote{https://skyview.gsfc.nasa.gov/current/cgi/titlepage.pl}.
As our target is corresponding nonthermal radio emission
from the possible SNR,
we also examined low frequency data
and found Sydney University Molonglo Sky Survey (SUMSS)
at 843~MHz \citep{mauch2003},
Murchinson Widefield Array 72--231~MHz \citep{hurleywalker2019},
and HI All-Sky Continuum Survey at 408~MHz \citep{haslam1982}.
In all these surveys we can find faint emission around 75~arcsec northeast of the pulsar position
as shown in Figure~\ref{fig:radio}.
Its spectral index is very steep, $< -1$.
However, 
strong contamination from a bright HII region prevents us from
the flux density and spectral index estimate.

There is no X-ray deep observation for this region.
Our 10~ks observation is too shallow to find diffuse X-ray emission
from the expected SNR.
Obiously we need follow-up observations in the X-ray band.

\section{Discussion}
\label{sec:discuss}

We have made the 3$\sigma$ tight flux upper-limit of
\\ \noindent
PSR~J1208$-$6238,
2.2--10.0$\times 10^{-14}$~erg~s$^{-1}$cm$^{-2}$ in the 0.5--7.0~keV band
(table~\ref{tab:point-UL}).
In this section, we discuss on the comparison with other high-B pulsars.
Here, we treat the power-model result with the typical absorption
($8\times 20^{21}$~cm$^{-2}$), 2.2$\times 10^{-14}$~erg~s$^{-1}$cm$^{-2}$
as ``typical upper-limit''
and the blackbody model result with the large absorption
($1.48\times 10^{22}$~cm$^{-2}$), 10.0$\times 10^{-14}$~erg~s$^{-1}$cm$^{-2}$
as ``conservative upper-limit''.
\citet{watanabe2019} catalogued X-ray properties of high-B pulsars.
To make comparison with this sample,
we converted the derived flux upper-limit for PSR~J1208$-$6238
to 0.3--10~keV band one,
to be 2.3--13$\times 10^{-14}$~erg~s$^{-1}$~cm$^{-2}$.
The luminosity upper limit becomes 2.6--14$\times 10^{31}$$d_3{}^2$~erg~s$^{-1}$,
where $d_3$ is the distance with the unit of 3~kpc.
This is quite small value compared with the spin down energy,
$1.5\times 10^{36}$~erg~s$^{-1}$.
The emission efficiency $\eta$,
which is conversion factor of spin-down energy to the X-ray luminosity
in the 0.3--10~keV band,
is less than $1.7\times 10^{-5}d_3{}^2$ for the typical case
and $9.3\times 10^{-5}d_3{}^2$ for the conservative case,
respectively.

There is rather large uncertainty on the upper-limit estimation of $\eta$.
The assumed parameters such as $\Gamma$ in the power-law models
and $kT$ in the blackbody models affect the results,
but Figure~\ref{fig:dependence} shows that
our assumption of $\Gamma = 2.0$ or $kT = 0.2$~keV makes
rather loose upper-limit.
The distance uncertainty also makes uncertainty of $\eta$.
However, the upper-limit of $\eta$ does not exceed $\sim 10^{-3}$
even with very large distance assumption such as 10~kpc.

In the case that the emission is blackbody,
we can estimate the upper-limit of the emitting region size.
It highly depends on the assumed absorption column,
temperature, and distance,
but the upper-limit of the radius of emitting region is
10~m to 130~km.
Considering the possibility that the emitting region is
only a part of the neutron star surface,
we could not reject any parameters.

\begin{figure}
\includegraphics[width=0.48\textwidth]{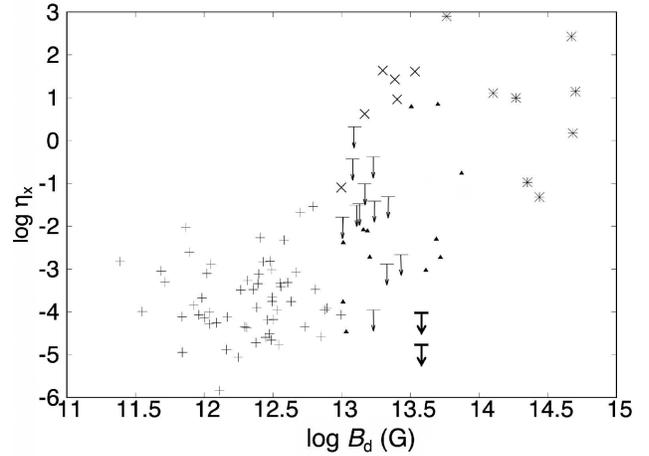}
\caption{$\eta$ vs. Dipole magnetic field for high-B pulsars.
Astrerisks, triangles, X-marks, and cross marks represent
magnetars, high-B pulsars, X-ray isolated neutron stars (XINSs),
and rotation-powered pulsars
from \citet{watanabe2019}.
Thick upper-limits are for typical (lower) and conservative (upper)
one for PSR~J1208$-$6238.
}

\label{fig:B-eta}
\end{figure}

Some high-B pulsars show burst activities similar to magnetars
(PSR~J1846$-$0258; \cite{kumar2008} PSR~J1119$-$6127; \cite{gogus2016,blumer2017}).
The period and period derivative are quite similar between
PSR~J1208$-$6238 and these two sources.
Figure~\ref{fig:B-eta} shows $\eta$ vs. the dipole magnetic field.
Magnetars have large X-ray luminosity and thus large $\eta$,
whereas conventional pulsars have $\eta$ of $10^{-3}$--$10^{-4}$.
One can see that PSR~J1208$-$6238 shows different properties
from those of magnetar-like high-B pulsar, PSR~J1846$-$0258 and PSR~J1119$-$6127;
or, considering the uncertainty of distance,
the small upper-limit of $\eta$ of PSR~J1208$-$6238 
is lower than typical value of those of high-B pulsars,
or even lower than those for samples with smaller magnetic field.
Our short exposure does not allow searching for magnetar-like flare activities.
Another interesting feature that needs further investigation is
searching for thermal radiation which is often seen in high-B pulsars
\citep{mclaughlin2007,zhu2011,ng2012,rigoselli2019}.
Deeper monitoring observations are also encouraged.

\begin{figure}
\includegraphics[width=0.47\textwidth]{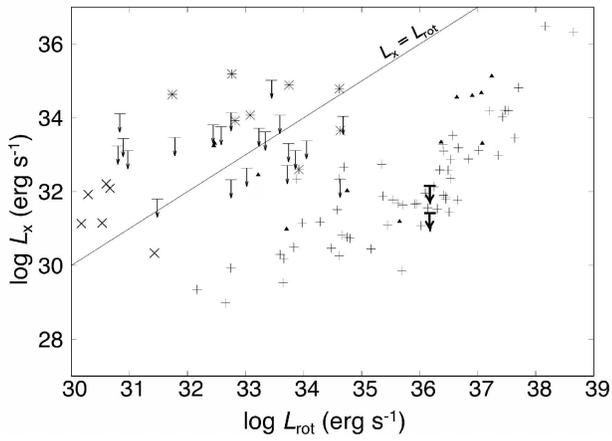}
\caption{Spin-down energy vs. 0.1--10~keV X-ray luminosity
of high-B pulsars.
Astrerisks, triangles, X-marks, and cross marks represent
magnetars, high-B pulsars, X-ray isolated neutron stars (XINSs),
and rotation-powered pulsars
from \citet{watanabe2019}.
Thick upper-limits are for typical (lower) and conservative (upper)
one for PSR~J1208$-$6238.
The upper-limit indicates 3$\sigma$ level for the all samples.
The line for the X-ray luminosity ($L_{\rm X}$) = the spin-down luminosity
($L_{\rm rot}$) is also shown.
}
\label{fig:efficiency}
\end{figure}

Figure~\ref{fig:efficiency} shows the summary of
X-ray luminosity in the 0.3--10~keV band vs. spin-down energy.
Typical spin-down powered pulsars have 
positive correlation between these two parameters
(\cite{shibata2016}; see also \cite{kargaltsev2008}),
whereas pulsars with magnetar-like activities have
brighter X-ray luminosity, $10^{35-36}$~erg~s$^{-1}$.
This figure shows that PSR~J1208$-$6238 has
very faint X-ray luminosity compared with 
other magnetars and spin-down powered pulsars.
It implies that PSR~J1208$-$6238 does not show
magnetar-like activities \citep{shibata2016}.
While we have large uncertainties of distance,
they does not significantly help 
magnetar-like origin.

PSR~J1208$-$6238 is a GeV gamma-ray emitting pulsar.
Interesting fact is that magnetars have not detected in the GeV band
so far,
which also imply that PSR~J1208$-$6238 is not a typical magnetar-like pulsar.
Non-detection in the X-ray band shows us that 
this source has a higher GeV to X-ray flux ratio
compared with other young rotation-powered pulsars
and similar to older samples \citep{cotizelati2020}.
This pulsar can be in the transition phase
from young ($\sim$ thousand years) to old system ($\sim$ a few thousand years).

We have found some signals from its possible PWN with 4.4$\sigma$ confidence level.
Assuming the distance of 3~kpc again,
the resultant luminosity 90\% upper-limit is 
4.3$\times 10^{31}$~erg~s$^{-1}$.
A similar pulsar, PSR~J1119$-$6127, have a faint PWN
with the luminosity of 
\\ \noindent
1.9$\times 10^{32}$~erg~s$^{-1}$
in the quiescent phase \citep{blumer2017},
suggesting that the possible PWN of PSR~J1208$-$6238 
can be similar to that of
\\ \noindent
PSR~J1119$-$6127.
In order to make clear the existence of this possible nebula,
deeper observations will be needed.

We found a nearby radio source in 77--231~MHz, 843~MHz and 408~MHz images.
As it is position somewhat further away from the high-B pulsar position
and it shows very steep spectral index, $< -1$,
which is totally different from those for typical SNRs,
we place low confidence 
to be considered as its associating SNR shock.
The nature of this emission is still unclear.
Since the pulsar is vey young, $\sim$2000~years old,
we also expect thermal and/or nonthermal X-rays
from the possible SNR.
Unfortunately, our exposure is too short to detect
such diffuse emission,
and no deep X-ray observation in this region has been performed.
\citet{clark2016} also searched for the host SNR.
The nearby gamma-ray emission shows soft spectrum like other gamma-ray SNRs,
but the distance between the emission and the pulsar
is too large to consider
the offset from the pulsar is due to the kick-off.
Deeper observations in multiwavelength
will be needed to study the surrounding environment
of this interesting pulsar.

\section{Summary}

X-ray follow-ups of high-B pulsars give us crucial information
on their subclass, such as the magnetars and the conventional rotation-powered pulsars.
We have made a 10~ks {\it Chandra} observation on
one of the youngest high-B pulsars
PSR~J1208$-$6238.
No significant X-ray emission was found on the pulsar region,
although there are some photon excess.
We derived the intrinsic flux and luminosity upper-limit of
2.2--10$\times 10^{-14}$~erg~s$^{-1}$cm$^{-2}$ in the 0.5--7.0~keV band
and 2.6--13$\times 10^{31}$$d_3{}^2$~erg~s$^{-1}$ in the 0.3--10~keV band,
respectively,
depending on the emission models.
The conversion efficiency of the spin-down energy to X-ray emission is
less than $1\times 10^{-4}$  with the typical spectral model,
which is quite small compared with magnetar-like high-B pulsars
or even conventional rotation-powered.
We also searched for the hosting supernova remnant
in the radio band and found no significant emission.

Deeper observations in multiwavelength are needed to follow-up possible flares
and putting deeper upper-limit.

\acknowledgments
We thank the anonymous referee 
for the productive comments.
We use the SkyView and Simbad virtual observatory.
This work is supported in part by 
Shiseido Female Researcher Science Grant (AB), 
Grant-in-Aid for Scientific Research of the Japanese
Ministry of Education, Culture, Sports, Science and Technology (MEXT) of Japan,
No.18H05459 (AB), 
No.19K03908 (AB), No.18H01246 (SS), No.16H03983 (KM),
and No.19K14758, 19H05075 (HS).

\end{document}